\documentclass[11pt,twoside,letterpaper]{article} 
\usepackage{times,fancyhdr}
\usepackage[dvips]{graphicx}


\makeatletter
\def\cleardoublepage{\clearpage\if@twoside \ifodd\c@page\else%
    \hbox{}%
    \thispagestyle{empty}%
    \newpage%
    \if@twocolumn\hbox{}\newpage\fi\fi\fi}
\makeatother

\def\figurename{Figure}
\makeatletter
\renewcommand{\fnum@figure}[1]{\figurename~\thefigure.}
\makeatother

\def\tablename{Table}
\makeatletter
\renewcommand{\fnum@table}[1]{\tablename~\thetable.}
\makeatother
\begin{document}

\title{
{\begin{flushleft}
\vskip 0.45in
{\normalsize\bfseries\textit{Chapter~4. In: Recent Advances in Cosmology, 2013 Nova Science Publishers, Inc., pp 97-127}}
\end{flushleft}
\vskip 0.45in
\bfseries\scshape Relativistic Viscous Universe Models}}
\author{\bfseries\itshape Iver Brevik$^{1}$  and  {\O}yvind Gr{\o}n$^{2}$ \\
$^{1}$ Department of Energy and Process Engineering,\\ Norwegian University of Science and Engineering, Trondheim, Norway\\
$^{2}$ Oslo and Akershus University College of Applied Sciences,\\ Faculty of Technology , Art and Design,  Oslo, Norway
}
\date{}
\maketitle
\thispagestyle{empty}
\setcounter{page}{1}

\begin{abstract}
  The research on relativistic universe models with viscous fluids is reviewed. Viscosity may have been of significance during the early inflationary era, and may also be of  importance for the late time evolution of the Universe. Bulk viscosity and shear viscosity cause exponential decay of anisotropy, while nonlinear viscosity causes power-law  decay of anisotropy. We consider also the influence from turbulence, in connection with future singularities of the universe (Big Rip and Little Rip). Finally, we review some recent developments of  causal cosmology theories.
\end{abstract}
\pagestyle{fancy}
\fancyhead{}
\fancyhead[EC]{Iver Brevik  and  {\O}yvind Gr{\o}n }
\fancyhead[EL,OR]{\thepage}
\fancyhead[OC]{Relativistic Viscous Universe Models}
\fancyfoot{}
\renewcommand\headrulewidth{0.5pt}

  \section{Viscous Universe Models}

   Misner [1967] \cite{misner67} noted that the 'measurement of the isotropy of the cosmic background radiation represents the most accurate observational datum in cosmology, which is even more true today with the WMAP- and Planck measurements. An explanation of this isotropy was provided by showing that in a large class of homogeneous but anisotropic universes, the anisotropy dies away rapidly. It was found that the most important mechanism in reducing the anisotropy is neutrino viscosity at temperatures just above $10^{10}$ K (when the Universe was about 1 s old: cf. Zel'dovich and Novikov [1983] \cite{zeldovich83}).
   The first theory of relativistic viscous fluid was presented in Eckart [1940] \cite{eckhart40}. Eckart's theory deals with first order deviation from equilibrium, while neglected second order terms are necessary to prevent non-causal behavior. Israel and Stewart [1976] \cite{israel76} have developed a second order theory. Gr{\o}n [1990] \cite{gron90} and Maartens [1995, 1996] \cite{maartens95,maartens96} have presented exhaustive reviews of research on cosmological models with non-causal and causal theories of viscous fluids, respectively.

   Bulk viscosity driven cosmic expansion with the Israel-Stewart theory have been investigated by Zimdahl [1996] \cite{zimdahl96}, Mak and Harko [1998] \cite{mak98}, Paul et al. [1998] \cite{paul98} and by Arbab and Beesham [2000] \cite{arbab00}. As noted by Lepe et al. [2008] \cite{lepe08}, although Eckart's theory presents some causality problems, it is the simplest alternative and has been widely considered in cosmology, as documented in Gr{\o}n [1990] \cite{gron90}, which we refer to for works on these topics up to 1990. We will here review papers from 1990 and onwards.

   Many types of observations favor that our universe is homogeneous and isotropic on scales above a billion light years. The observations of the temperature fluctuations in the cosmic microwave radiation favor that the universe is flat, i. e. that the total density of the matter and energy contained in the universe is equal to the critical density.

   The discovery that the expansion of the universe accelerates could most simply be explained by repulsive gravity due to a cosmic vacuum energy with a density equal to about 70\% of the critical density. The observations also favor a special type of vacuum energy which may be represented by a cosmological constant in Einstein's field equations. The energy-momentum tensor of this energy is proportional to the metric tensor. One may show that this means that every component of the energy-momentum tensor is Lorentz invariant  [Zel'dovich 1968] \cite{zeldovich68} and Gr{\o}n [1986] \cite{gron85}. Hence it is not possible to measure velocity with respect to this type of energy. It may therefore be called a Lorentz invariant vacuum energy, LIVE.

   Furthermore a large amount of cold dark matter is needed to keep the galaxies and the hoops of galaxies together because of the rapid motions of the stars in the galaxies and of the galaxies in the hoops. Hence, about 30 \% of the contents of the universe seem to be in the form of cold dark matter.
   The cosmologists therefore introduced a standard model of the universe dominated by two fluids, a Lorentz invariant vacuum energy, LIVE, and a cold fluid. The vacuum energy is usually called {\it dark energy} and the cold fluid is called {\it dark matter}. Since the observations of the temperature fluctuations in the cosmic microwave radiation favor that the universe is flat, we shall only consider flat universe models.

   In this review we will focus upon universe models based upon the general theory of relativity. This means among other things that we shall only review works in which the gravitational parameter is constant. It should be mentioned, though, that some researchers have investigated universe models with variable gravitational parameter (see, for instance, Belinch\'{o}n, Harko and Mak [2002] \cite{belinchon02}).

\section{The Standard Model of the Universe}

For later comparison we shall first briefly summarize the main properties the standard model which has vanishing viscosity. In this model the total pressure and density are given by	
\begin{equation}
\rho=\rho_M+\rho_\Lambda, \quad p=p_M+p_\Lambda=-\rho_\Lambda, \label{2.1}
\end{equation}
                                                           where $\rho_M$  is the density of cold dark matter in the form of dust, $p_M=0$,  and  $\rho_\Lambda$ is the density of a Lorentz Invariant Vacuum Energy, LIVE, which has $p_\Lambda=-\rho_\Lambda $  (using units so that $c = 1 $), and may thus be represented by a cosmological constant in Einstein's field equations.

The line-element has the form (using units so that the velocity of light in empty space is equal to 1),
\begin{equation}
ds^2=-dt^2+a(t)^2(dr^2+r^2d\Omega^2), \quad d\Omega^2=d\theta^2+\sin^2\theta d\phi^2. \label{2.2}
\end{equation}
                                                                       The scale factor of this model is [Gr{\o}n 2002] \cite{gron02}
\begin{equation}
 a=K_s^{1/3} \sinh^{2/3}(\frac{t}{t_\Lambda}),
  \quad t_\Lambda=\frac{2}{\sqrt{3\kappa \rho_\Lambda}}
 =\frac{2}{3H_0 \sqrt{\Omega_{\Lambda 0}}}, \quad K_s=\frac{1-\Omega_{\Lambda 0}}{\Omega_{\Lambda 0}}, \label{2.3}
 \end{equation}
where  $\kappa=8\pi G$ is Einstein's gravitational constant. Here  $H_0$ and $\Omega_{\Lambda 0}$  are the present values of the Hubble parameter and the density parameter of LIVE. The scale factor represents the distance between two galaxy clusters relative to their present distance. Hence $a(t_0)=1$  where the present age of the universe is
                \begin{equation}
                t_0=t_\Lambda \,{\rm arctanh}
                \sqrt {\Omega_{\Lambda 0}}.  \label{2.4}
                \end{equation}
                Inserting the presently favored values $13.7\times 10^9$  years and $\Omega_{\Lambda 0}=0.7$ leads to $t_\Lambda=11.4\times 10^9$   years.

The Hubble parameter is
\begin{equation}H=(2/3t_\Lambda)\coth (t/t_\lambda). \label{2.5}
\end{equation}

The deceleration parameter is
\begin{equation}
q=-\frac{a\ddot{a}}{\dot{a}^2}=-1-\frac{\dot{H}}{H^2}, \label{2.6}
\end{equation}
giving
\begin{equation}
q=(1/2)[1-3\tanh^2(t/t_\Lambda)]. \label{2.7}
\end{equation}

\begin{figure}[htb]
  \centerline{\includegraphics[width=0.8\textwidth]{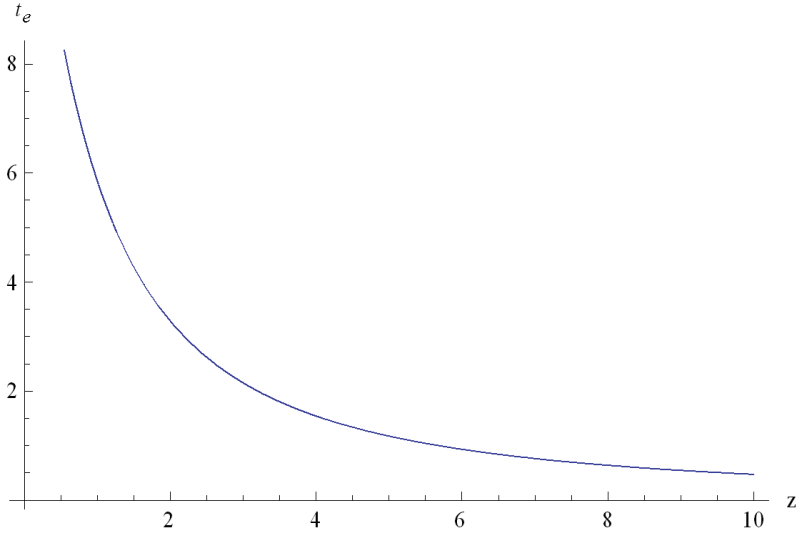}}
  \caption{Emitter time in billion years as a function of cosmic redshift.}
  \label{fig1}
\end{figure}

           The point of time $t_1$  when deceleration turns into acceleration is given by  $q(t_1)=0$ which leads to
           \begin{equation}
           t_1=t_\Lambda {\rm arctanh} (1/\sqrt 3). \label{2.8}
           \end{equation}
The corresponding redshift is
\begin{equation}
z= \frac{1}{a(t_1)}-1=\left(\frac{2\Omega_{\Lambda 0}}{1-\Omega_{\Lambda 0}}\right)^{1/3}-1, \label{2.9}
\end{equation}
which gives $t_1=7.4\times 10^9$  years and $z(t_1)=0.67$.
In this model the age-redshift relationship is
\begin{equation}
t_e=t_0 \,\frac{{\rm arcsinh} \frac{\sqrt{\Omega_{\Lambda 0}}}{\sqrt{\Omega_{M0}}(1+z)^{3/2}}}
{{\rm arcsinh}\sqrt{\Omega_{\Lambda 0}/\Omega_{M0}}}. \label{2.10}
\end{equation}
Here $t_e$ is the emitter time of a signal arriving at the instant $t_0$. With a unit of time equal to $10^9$ years and by inserting $t_0=13.7$, $\Omega_{\Lambda 0}=0.7$ we can write equation (\ref{2.10}) as
\begin{equation}
t_e=11.3\, {\rm arcsinh} \left[ 1.53 (1+z)^{-1.5}\right]. \label{2.11}
\end{equation}
This is shown graphically in Figure \ref{fig1}.

\newpage

\setcounter{equation}{0}
\section{	Viscous Fluid in an Expanding Universe}

   Let  $u^\mu$ be components of the 4-velocity of a fluid element. The projection tensor onto a 3-space orthogonal to the world line of a fluid element is defined by
   \begin{equation}
   h_{\alpha\beta}=g_{\alpha\beta}+u_\alpha u_\beta. \label{3.1}
   \end{equation}

The covariant derivative of the velocity field of the fluid can be written as a 3 times 3 matrix which can be separated into the antisymmetric part representing the {\it vorticity},
\begin{equation}
\omega_{\alpha\beta}=\frac{1}{2}(u_{\mu;\nu}-u_{\nu;\mu})h_\alpha^\mu h_\beta^\nu, \label{3.2}
\end{equation}
the trace free, symmetrical part which represents the {\it shear},
\begin{equation}
\sigma_{\alpha\beta}=\left[ \frac{1}{2}(u_{\mu;\nu}+u_{\nu;\mu})-\frac{1}{3}u_{;\lambda}^\lambda  h_{\mu\nu}\right] h_\alpha^\mu h_\beta^\nu, \label{3.3}
\end{equation}
and the  trace, which represents the {\it expansion},
\begin{equation}
\theta= u_{;\lambda}^\lambda. \label{3.4}
\end{equation}

The four-acceleration of the fluid element, which is non-vanishing only for non-geodesic flow, is defined by
\begin{equation}
a_\alpha=a_{\alpha;\mu}u^\mu. \label{3.5}
\end{equation}
We then have
\begin{equation}
u_{\alpha;\beta}=\omega_{\alpha;\beta}+\sigma_{\alpha;\beta}+\frac{1}{3}\theta h_{\alpha\beta}-\sigma_\alpha u_\beta. \label{3.6}
\end{equation}

   The energy-momentum tensor of a viscous fluid with proper density $\rho$  and pressure $p$ is
   \begin{equation}T_{\alpha\beta}=\rho u_\alpha u_\beta +(p-\xi \theta)h_{\alpha\beta}-2\eta \sigma_{\alpha\beta}, \label{3.7}
   \end{equation}
where $\eta$ and  $\xi$  are the coefficients of shear and bulk viscosity, respectively. Einstein's field equations imply that the divergence of this tensor vanishes. From this one may deduce the equation of continuity of the fluid in the form [Kohli 2012] \cite{kohli12}
\begin{equation}
\dot{\rho}+(\rho +p)\theta-4\eta \sigma^2-\xi \theta^2=0. \label{3.8}
\end{equation}

 The evolution of the divergence with time is given by the Raychaudhuri equation which may be written as [Ellis 2007] \cite{ellis07}
 \begin{equation}
 \dot{\theta}= a_{;\lambda}^{\lambda}+\frac{\kappa}{2}(3\xi \theta -\rho-3p)+2\omega^2-2\sigma^2-\frac{1}{3}\theta^2+\Lambda. \label{3.9}
\end{equation}

   For a universe model containing a fluid with equation of state  $p_M=w \rho_M$ and a LIVE interacting with each other, the interaction is usually represented by a term  $Q(t)$ so that the equations of continuity of the fluids take the form [Jamil and Farooq [2010] \cite{jamil10}
   \begin{equation}
   \dot{\rho}_M+(1+w)\rho_M\theta =-Q, \quad \dot{\rho}_\Lambda=Q. \label{3.10}
   \end{equation}

\section{	Isotropic, Viscous Generalization of the Standard Universe Model}
\setcounter{equation}{0}

   We now consider a homogeneous and isotropic universe with geodesic fluid flow. In this case  $a_{;\lambda}^\lambda=\omega=\sigma=0$  and eqs. (\ref{3.8}) and (\ref{3.9}) in Sect. 3 reduce to
   \begin{equation}
   \dot{\rho}+(\rho+p)\theta-\xi \theta^2=0, \label{4.1}
   \end{equation}
   and
   \begin{equation}
   \dot{\theta}=\frac{\kappa}{2}(3\xi \theta-\rho-3p)-\frac{1}{3}\theta^2+\Lambda. \label{4.2}
   \end{equation}

Assuming that the 3-space is Euclidean the line element takes the form
\begin{equation}
ds^2=-dt^2+a(t)^2(dr^2+r^2d\theta^2+r^2\sin^2\theta d\phi^2). \label{4.3}
\end{equation}

With co-moving coordinates in the cosmic fluid the expansion of the fluid and the Hubble parameter are related by $\theta=3H$. We shall assume that the universe contains two non-interacting fluids, Lorentz invariant vacuum energy, LIVE, that may be represented by a cosmological constant, $\kappa \rho_\Lambda=\Lambda$,  and a fluid with density $\rho_M$ and pressure  $p_M=w\rho_M$. In this case Einstein's field equations take the form
\begin{equation}
 \kappa \rho_M=3H^2-\Lambda, \label{4.4}
\end{equation}
and
\begin{equation}
\frac{\ddot{a}}{a}=-\frac{\kappa}{6}(1+3w)\rho_M+\frac{\Lambda}{3}+\frac{3\kappa \xi}{2}H. \label{4.5}
\end{equation}

Using that
\begin{equation}
\frac{\ddot{a}}{a}=\dot{H}+H^2, \label{4.6}
\end{equation}
these equations lead to
\begin{equation}
\dot{H}=-\frac{3}{2}(1+w)H^2+\frac{3\kappa \xi}{2}H+\frac{1}{2}(1+w)\Lambda, \label{4.7}
\end{equation}                                                                                                                  in accordance with eq.(\ref{4.2}).  The critical density is given by  $\kappa \rho_{cr}=3H^2$.  The density parameters of the fluid and the LIVE are $\Omega_M=\rho_M/\rho_{cr}, \, \Omega_\Lambda=\Lambda/\kappa \rho_{cr}$.   Defining a corresponding viscosity parameter $\Omega_{\xi 0}=\kappa \xi_0/H_0$  and using the definition (\ref{2.6}) in Sect. 2 of the deceleration parameter equation (\ref{4.7}) for the present  time can be written as
\begin{equation}
q_0=\frac{1}{2}(1+3w)-\frac{3}{2}[\Omega_{\xi 0}+(1+w)\Omega_{\Lambda 0}]. \label{4.8}
\end{equation}

Hence, the present value of the viscosity parameter of the mixture of a cosmic fluid and LIVE can be given in terms of the present values of the Hubble parameter, the deceleration parameter and the density parameter of LIVE. If the cosmic fluid is cold we may put $w=0$, which leads to
\begin{equation}
\Omega_{\xi 0}=\frac{1}{3}(1-2q_0)-\Omega_{\Lambda 0}.\label{4.9}
\end{equation}
Pavon and Zimdahl [1993] \cite{pavon93} have used a corresponding formula to estimate the value of  $\Omega_{\xi 0}$ that is necessary in order that the viscosity of the cosmic fluid shall be of significance for the evolution of the expansion history of the universe.  A recent determination of the present value of the deceleration parameter from observational data give large uncertainties [Giostri et al 2012] \cite{giostri12},  $-0.66 <q_0<-0.20$. Furthermore, $0.69<\Omega_{\Lambda 0}<0.77.$  Inserting these values into eq.(\ref{4.9}) gives a lower boundary on  $\Omega_{\xi 0}$  which is negative, which means that these data do not require any viscosity at all.  Hence the motivation for studying cosmological models with viscous fluids does not come from observational data, but from the possibility of the existence of physical mechanisms that can produce viscosity. One such possibility that has been studied by Mathews et al [2008] \cite{mathews08}, is production of viscosity by decay
 of dark matter particles to relativistic particles at a recent epoch with redshift  $z<1$.

Before reviewing more general models we shall consider the most simple viscous universe model. It is dominate by dust with constant viscosity coefficient, and has been studied by Padmanabhan and Chitre [1987] \cite{padmanabhan87}. This model is reviewed not because it is physically realistic, but because it gives us an opportunity to demonstrate in a simple way typical features of universe models dominated by viscous fluids. For this model eq. (\ref{4.7}) reduces to
\begin{equation}
\dot{H}=-\frac{3}{2}H^2+\frac{3}{2}\Omega_{\xi 0}H_0H. \label{4.10}
\end{equation}
Integrating this equation two times with the boundary conditions $H(t_0)=H_0, a(0)=0, a(t_0)=1$ leads to
\begin{equation}
H=\frac{1}{2}\Omega_{\xi 0}H_0\left[ 1+\coth \left( \frac{3}{4}\Omega_{\xi 0}H_0t\right)\right],
 a=\left[ \frac{1-\Omega_{\xi 0}}{\Omega_{\xi 0}}\right]^{2/3}
\left( e^{\frac{3}{2}\Omega_{\xi 0}H_0t}-1\right)^{2/3}, \label{4.11}
\end{equation}
where the age of the universe is given in terms of the present value of the Hubble parameter as
\begin{equation}
t_0=\frac{4}{3\Omega_{\xi 0}H_0}{\rm arctanh}\frac{\Omega_{\xi 0}}{2-\Omega_{\xi 0}}. \label{4.12}
\end{equation}
If $\Omega_{\xi 0}H_0t \ll 1$ the scale factor takes the approximate form $a \approx [(3/2)H_0t]^{2/3}$ corresponding to the evolution of the universe dominated by viscosity free dust. Hence in this universe model the viscosity can be neglected at early times. On the other hand at late times so that $\Omega_{\xi 0} H_0t \gg1$ the expansion becomes exponential with $H=\kappa \xi_0, a\propto \exp(\kappa \xi_0), \rho =3\kappa \xi_0^2$, and the universe enters a late inflationary era with accelerated expansion. This was pointed out by Panmanabhan and Chitre [1987] \cite{padmanabhan87} more than ten years before the late accelerated expansion of the universe was discovered. However, the time $t_c$ after the viscosity becomes dominant may be be extremely late if there is no effective mechanism for creating viscosity in the cosmic fluids,
\begin{equation}
t_c=\frac{2}{3\Omega_{\xi 0}H_0}=\frac{c^2}{12\pi G\xi_0}=1.5 \times 10^{10} {\rm years}
 \left(\frac{10^9 { {\rm g  cm} ^{-1}{\rm s}^{-1}}}{\xi_0}\right). \label{4.13}
\end{equation}

   Ren and Meng [2006] \cite{ren06},  Hu and Meng [2006] \cite{hu06}, and Mostafapoor and Gr{\o}n [2011] \cite{mostafapoor11} have studied universe models in which the bulk viscosity coefficient of the viscous fluid has the
    following form
   \begin{equation}
   \xi=\xi_0+\xi_1\frac{\dot a}{a}+\xi_2\frac{\ddot a}{\dot a}\,. \label{4.14}
  \end{equation}

The motivation for considering this form for the coefficient of bulk viscosity is that from fluid mechanics we know that the viscosity is related to the motion of the fluid, i.e. to  $\dot a$ and $\ddot a$.    Inserting eq.(4.7) into eq.(4.6) we obtain
\begin{equation}
a\dot{H}=-bH^2+cH+d, \label{4.15}
\end{equation}
where
\begin{equation}
a=1-\frac{3\kappa \xi_2}{2}, \quad b=\frac{3}{2}[1+w-\kappa (\xi_1+\xi_2)], \quad c=\frac{3\kappa \xi_0}{2}, \quad d=\frac{1}{2}(1+w)\Lambda. \label{4.16}
\end{equation}

   Integration with  $a(0)=0, \, a(t_0)=1$  and assuming  $\kappa(\xi_1+\xi_2)<1$ and $w \geq 0$ so that $b>0$ and $4bd+c^2>0$ gives
   \begin{equation}
   H(t)=\frac{c}{2b}+\frac{a}{b}\hat{H}\coth (\hat{H}t), \quad \hat{H}^2=\frac{bd}{a^2}+\frac{c^2}{4a^2}, \label{4.17}
   \end{equation}
and
\begin{equation}
a(t)=e^{\frac{c}{2b}(t-t_0)}\left[ \frac{\sinh (\hat{H}t)}{\sinh (\hat{H}t_0)}\right]^{a/b}, \label{4.18}
\end{equation}

or
\begin{equation}
a(t)=(K_\xi)^{\frac{a}{2b}}e^{\frac{c}{2b}(t-t_0)}\sinh^{\frac{a}{b}}
(\hat{H}t), \quad K_\xi=\frac{4b(bH_0^2-cH_0-d)}{4bd+c^2 }, \label{4.19}
\end{equation}
where
\begin{equation}
t_0=\frac{1}{\hat{H}}{\rm arctanh} \frac{2a\hat{H}}{2bH_0-c} \label{4.20}
\end{equation}
is the age of the universe. The corresponding scale factor and age of the universe if the viscosity vanishes are given by equations (\ref{2.3}) and (\ref{2.4}) in Sect. 2. It follows that for a given present value of the Hubble parameter the viscosity increases the age of the universe. Assuming that  $\kappa \xi_0 \ll H_0$  the increase of the age due to the viscosity is of an order of magnitude
\begin{equation}
t_0-t_{00} \approx \Omega_{\xi 0}^2 \,t_0. \label{4.21}
\end{equation}

Brevik and Heen [1994] \cite{brevik94} have used that in the plasma era of the universe the bulk viscosity derived from kinetic theory of gases has order of magnitude so that $(\kappa \xi_0)^{-1} \sim 10^{21}$ years. Since  $(H_0)^{-1} \sim 10^{10}$ years this estimate of the magnitude of the bulk viscosity gives  $\Omega_{\xi 0} \sim 10^{-11}$. During most of the evolution of the universe the viscosity is smaller than this. Hence, this form of viscosity is totally insignificant for the age of the universe.

   Brevik and Stokkan [1996] \cite{brevik96} have, however, pointed out that impulsive processes at the end of the inflationary era may have produced great viscosity. In this extremely brief period the viscosity may have given significant contributions to the production of entropy in the universe, able to explain why the number of photons per baryon is so large, $\sim 10^9$, in our universe.

   With vanishing cosmological constant, i.e. $d=0$, the relationship between the Hubble parameter and the cosmological redshift  $1+z=a^{-1}$ is
   \begin{equation}
   H(z)=\frac{c}{b}+\left(H_0-\frac{c}{b}\right)(1+z^{b/a}). \label{4.22}
   \end{equation}

Dou and Meng [2011] \cite{dou11} have considered a universe model with $w=\xi_1=\xi_2=0$.  Then  $c/b=\Omega_{\xi 0}H_0, \, b/a=3/2$  and eq.(\ref{4.22}) reduces to
\begin{equation}
H(z)=H_0\left[ \Omega_{\xi 0}+(1-\Omega_{\xi 0})(1+z)^{3/2}\right]. \label{4.23}
\end{equation}
   Avelino and Nucamendi [2008] \cite{avelino08}  have used supernova data to constrain the dimensionless viscosity parameter  $\Omega_{\xi 1}=\kappa \xi_1$  by considering a matter-dominated universe model with bulk viscosity proportional to the Hubble parameter. In this case  $\xi_0=\xi_2=w=\Lambda=0$ giving  $a=1, \, c=d=0$, and eq. (\ref{4.15}) reduces to
   \begin{equation}
   \dot{H}=-bH^2. \label{4.24}
   \end{equation}

The Hubble parameter, scale factor and density then are
\begin{equation}
H=\frac{H_0}{1+bH_0(t-t_0)}, \quad a=\frac{1}{[1+bH_0(t-t_0)]^{1/b}}, \quad \rho=\frac{\rho_0}{[1+bH_0(t-t_0)]^2}, \label{4.25}
\end{equation}
with  $b=(3/2)(1-\Omega_{\xi 1})$.  In terms of the redshift z the Hubble parameter can in this case be expressed as
\begin{equation}
H(z)=H_0(1+z)^{(3/2)(1-\Omega_{\xi 1})}. \label{4.26}
\end{equation}

Using a combination of observational data including emitter points of time at the early universe, i.e. with  $z \gg 2$, Avelino and Nucamendi [2008] \cite{avelino09} found that  $\Omega_{\xi 1}$ had to be negative in order that the properties of this universe model should not be in conflict with the data. But using only supernova data with $z<2$  they found that agreement with the observations requires $\Omega_{\xi 1}=0.48 \pm 0.04$,  which is not physically impossible. Hence, they concluded that in order that a matter-dominated universe model with bulk viscosity proportional to the Hubble parameter shall be in agreement with observations, the viscosity must be due to a mechanism that produces viscosity mainly in the recent history of the universe. This is in agreement with the conclusion of Mathews et al [2008] \cite{mathews08}.

   From a general form of the solution corresponding to eq.(\ref{4.17}) applied to the case of a universe with no dark matter, only dominated by dark energy with  $w=-1$  with linear viscosity, i.e.   $\xi_1=\xi_2=0$ so that  $b=0$, Cataldo, Cruz and Lope [2005] \cite{cataldo05} concluded that the scale factor is not defined by the field equations. However, this form of the solution is only valid for  $b \neq 0$. The field equations determine the scale factor also in this case where eq.(\ref{4.7}) reduces to
   \begin{equation}
   \dot{H}=\frac{3\kappa \xi_0}{2}H. \label{4.27}
   \end{equation}

Integration with $a(t_0)=0$ gives
\[
H(t)=H_0 \exp\left[\frac{3\Omega_{\xi 0}H_0}{2}(t-t_0)\right], \]
\begin{equation}
 a(t)=\exp\left\{ \frac{2}{3\Omega_{\xi 0}}\left[ e^{\frac{3\Omega_{\xi 0}H_0}{2}(t-t_0)}-1\right]\right\}.  \label{4.28}
\end{equation}
Hence a universe dominated by viscous dark energy with constant viscosity coefficient expands exponentially faster than a corresponding universe model without viscosity.

  The form of the solution when $4bd+c^2<0$  is
  \begin{equation}
  H(t)=\frac{c}{2b}+\frac{a}{b}\hat{H}\cot (\hat{H}t), \quad \hat{H}^2=-\left(\frac{bd}{a^2}+\frac{c^2}{4a^2}\right), \label{4.29}
  \end{equation}
and
\begin{equation}
a(t)=e^{\frac{c}{2b}(t-t_0)}\left[ \frac{\sin(\hat{H}t)}{\sin(\hat{H}t_0)}\right]^{a/b}. \label{4.30}
\end{equation}

Note that this solution is only valid for $d\neq 0$.  The scale factor blows up to infinity, i.e. there is a Big Rip, at a point of time $t_R$   given by $\hat{H}t_R=\pi$,  or
\begin{equation}
t_R=\frac{2a\pi}{\sqrt{-(4bd+c^2)}}. \label{4.31}
\end{equation}

\section{	Viscosity and the Accelerated Expansion of the Universe}
\setcounter{equation}{0}

   The question whether a matter dominated universe with bulk viscosity can drive the accelerated expansion of the universe has been discussed by Kremer and Devecchi [2003] \cite{kremer03},  Fabris, Goncalves and Ribeiro [2006] \cite{fabris06},  Avelino and Nucamendi [2009] \cite{avelino09}. In the universe model of Avelino and Nucamendi  $\xi_1=\xi_2=0, \, w=0, \, \Omega_M=1, \, \Omega_\Lambda=0$,   and the expression (\ref{4.12}) in Sect. 4 for the scale factor reduces to
   \begin{equation}
   a(t)=\left[ \frac{4(1-\Omega_{\xi 0})}{\Omega_{\xi 0}^2}\right]^{1/3}
   \exp \left[(\Omega_{\xi 0}/2)H_0(t-t_0)\right]
    {\rm sinh}^{\frac{2}{3}}\left( \frac{3}{4}\Omega_{\xi 0}H_0 t\right), \label{5.1}
   \end{equation}
where the age of the universe is
\begin{equation}
t_0=\frac{4}{3\Omega_{\xi 0}H_0}{\rm arctanh} \frac{\Omega_{\xi 0}}{2-\Omega_{\xi 0}}=-\frac{2}{3\Omega_{\xi 0}H_0}\ln (1-\Omega_{\xi 0}). \label{5.2}
\end{equation}

This form of the solution satisfies the boundary conditions  $a(0)=0, a(t_0)=1$. The first of these are not satisfied by the form of the solution given by Avelino and Nucamendi. The solution (\ref{5.1}) may be written as
\begin{equation}
a(t)=\left(\frac{1-\Omega_{\xi 0}}{\Omega_{\xi 0}}\right)^{2/3}\left(e^{\frac{3}{2}\Omega_{\xi 0}H_0t}-1\right)^{2/3}. \label{5.3}
\end{equation}
This universe model has earlier been considered by Brevik and Gorbunova [2005] \cite{brevik05} and by Gr{\o}n [2010] \cite{gron10}, and is also identical to the model in eq.~(\ref{4.11}) in Sect. 4. The Hubble parameter is
\begin{equation}
H(t)=\frac{\Omega_{\xi 0}H_0}{1-e^{-(3/2)\Omega_{\xi 0}H_0t}}. \label{5.4}
\end{equation}
 It approaches the De Sitter model for $t \gg 1/\Omega_{\xi 0}H_0$ with a constant Hubble parameter equal to $\Omega_{\xi 0}H_0$. The deceleration parameter  is
\begin{equation}
q=\frac{3}{2\exp[(3/2)\Omega_{\xi 0} H_0 t]}-1, \label{5.5}
\end{equation}
with present value
\begin{equation}
q(t_0)=(1/2)(1-3\Omega_{\xi 0}). \label{5.6}
\end{equation}
These expressions show that the expansion starts from a Big Bang with an infinitely great expansion velocity, but decelerates to a finite value. At the instant $t_1$ given by $q(t_1)=0$ there is a transition to accelerated expansion, which will last for ever. The transition happens at
\begin{equation}
t_1=\frac{2\ln (3/2)}{3\Omega_{\xi 0}H_0}. \label{5.7}
\end{equation}
At this instant the scale factor has the value
\begin{equation}
a(t_1)=\left(\frac{1-\Omega_{\xi 0}}{2\Omega_{\xi 0}}\right)^{2/3}. \label{5.8}
\end{equation}
The corresponding redshift is
\begin{equation}
z_1=\left( \frac{2\Omega_{\xi 0}}{1-\Omega_{\xi 0}}\right)^{2/3}-1. \label{5.9}
\end{equation}
In order that the transition shall have happened at a past time, $a(t_1)<1$, the bulk viscosity must be sufficiently large, $\Omega_{\xi 0} > 1/3$.

For this universe model, with Euclidean spatial geometry, the matter density is equal to the critical density,
\begin{equation}
\rho_M=\frac{3H^2}{\kappa}=\frac{3\Omega_{\xi 0}^2H_0^2}{\kappa \left( 1-e^{-(3/2)\Omega_{\xi 0}H_0 t}\right)^2}. \label{5.10}
\end{equation}
Hence, the matter density approaches a constant value, $\rho_M \rightarrow (3/\kappa)\Omega_{\xi 0}^2 H_0^2$.

Avelino and Nucamendi [2009] \cite{avelino09} have used the most comprehensive supernova data to estimate the value of $\Omega_{\xi 0}$ that gives the best fit with observed data for a universe model containing dust with constant viscosity coefficient. The result was $\Omega_{\xi 0}=0.64$, which is eleven orders of magnitude greater than the value coming from kinetic gas theory \cite{brevik94}. However, a mechanism for producing greater viscosity may be  a bulk viscosity generation due to decay of dark matter particles into relativistic products; cf. Singh   [2008] \cite{singh08B}.

The final fate of a universe dominated by a viscous fluid has been discussed by Brevik and Gorbunova  [2005] \cite{brevik05} and by Cataldo, Cruz and Lepe  [2005] \cite{cataldo05}. Consider first a universe without viscosity and dark energy, containing only a non-viscous fluid with $p=w\rho$. In this case eq. (\ref{4.15}) in Sect. 4  reduces to
\begin{equation}
\dot{H}=-bH^2, \quad b=\frac{3}{2}(1+w). \label{5.11}
\end{equation}
The Hubble parameter, scale factor and density are given by eq. (\ref{4.15}) in Sect. 4 with $b$ given by eq. (\ref{5.11}). For this universe model there is a Big Rip at an instant
\begin{equation}
t_{R0}=t_0+\frac{2}{3(1+w)H_0}. \label{5.12}
\end{equation}
Cataldo et al. then considered a universe model with a fluid having $w<-1$ and constant viscosity coefficient $\xi_0$. Then $b<0$ and $d=0$ so neither of the above solutions are valid. In this case eq. (\ref{4.15}) in Sect. 4 reduces to
\begin{equation}
\dot{H}=-\frac{3}{2}(1+w)H^2+\frac{3}{2} \Omega_{\xi 0}H_0 H. \label{5.13}
\end{equation}
The Hubble parameter, scale factor and density for this universe model are
\begin{equation}
H=\frac{H_0}{\frac{1+w}{\Omega_{\xi 0}}+\left( 1-\frac{1+w}{\Omega_{\xi 0}}\right)e^{-\frac{3}{2}\Omega_{\xi 0}(t-t_0)}}, \label{5.14}
\end{equation}
\begin{equation}
a=\left[ 1-\frac{1+w}{\Omega_{\xi 0}}+\frac{1+w}{\Omega_{\xi_0}}e^{\frac{3}{2}\Omega_{\xi 0}H_0(t-t_0)}\right]^{\frac{2}{3(1+w)}}, \label{5.15}
\end{equation}
and
\begin{equation}
\rho=\frac{\rho_0}{\left[ \frac{1+w}{\Omega_{\xi 0}}+\left( 1-\frac{1+w}{\Omega_{\xi 0}}\right)e^{-\frac{3}{2}\Omega_{\xi 0}(t-t_0)}\right]^2}. \label{5.16}
\end{equation}
In this case there is a Big Rip at the instant
\begin{equation}
t_R=t_0+\frac{2}{3\Omega_{\xi 0}H_0}\ln \left( 1-\frac{\Omega_{\xi 0}}{1+w}\right). \label{5.17}
\end{equation}
To second order in $\Omega_{\xi 0}/(1+w)$ this formula gives
\begin{equation}
t_R \approx t_{R0}-\frac{\Omega_{\xi 0}}{3(1+w)^2H_0}, \label{5.18}
\end{equation}
where $t_{R0}$ is given in eq. (\ref{5.12}). Hence the viscosity makes the Big Rip come earlier.

Cataldo et al. [2005] \cite{cataldo05} have also considered a universe model in which the viscosity coefficient of the fluid is proportional to the square root of the density, $\xi =\alpha \rho^{1/2}$. Since the mass density is equal to the critical density, $\kappa \rho =3H^2$, this corresponds to the case (\ref{4.19}) with $b=(3/2)(1+w-\sqrt{3}\alpha)$, where $\kappa$ has been absorbed into $\alpha$. This again is equivalent to a universe model with only one viscous fluid where the coefficient of viscosity is proportional to the Hubble parameter.

Similar universe models with variable gravitational and cosmological 'constants' have been investigated by Singh, Beesham and Mbokazi [1998] \cite{singh98} and by Singh, Kumar and Pradhan [2007] \cite{singh07}.

\section{Viscous Bianchi Type-I Universe Models}
\setcounter{equation}{0}

The influence of viscosity on Bianchi type-I universe models, which are the anisotropic generalizations of the flat Friedmann-Robertson-Walker universe models, has been investigated by Belinski and Khalatnikov [1975] \cite{belinski75}. They found that for large times such universes models with constant coefficients of bulk viscosity, will approach an isotropic steady-state universe model with de Sitter spacetime which expands exponentially. For asymptotically early times they found that there exists a Kasner era in which the effect of matter, radiation and viscosity are negligible. Heller [1978] \cite{heller78} has concluded that such universe models have in general a stage near an unavoidable initial singularity in which the energy-momentum tensor has no influence on the cosmic evolution. However, Gr{\o}n [1990] \cite{gron90} has found that in a Bianchi type-I universe filled with viscous Zel'dovich fluid the bulk viscosity may remove the initial singularity.

In this section we will review articles where the influence of viscosity on the evolution of anisotropic Bianchi universe models filled with viscous fluids has been studied, following Mostafapoor and Gr{\o}n [2012] \cite{mostafapoor12}.

The line element of a Bianchi type-I universe can be written in the form
\begin{equation}
ds^2= dt^2-R_i^2\cdot (dx^i)^2, \label{6.1}
\end{equation}
where $R_1=a(t), R_2=b(t), R_3=c(t)$ are directional scale factors. The energy-momentum tensor of the viscous fluid has the non-vanishing components
\begin{equation}
T_0^0=\rho, \quad T_i^i=-p+2\eta H_i+(3\xi -2\eta)H-9\alpha H\Delta H, \label{6.2}
\end{equation}
where $ H_i=\dot{R}_i/R_i$ are the directional Hubble parameters, $H=(1/3)\sum_{i=1}^3 H_i, \,$ \mbox{$\Delta H_i=H_i-H$},  $p=w\rho$, and $\alpha$ is a non-linear viscosity coefficient. For these universe  models the Raychaudhuri equation takes the form
\begin{equation}
\dot{H}=-3H^2+\frac{\kappa}{2}(1-w)\rho+\frac{3}{2}\kappa \xi H+\Lambda. \label{6.3}
\end{equation}
The anisotropy parameter is defined as \cite{gron85}
\begin{equation}
A=\frac{1}{3}\sum_{i=1}^3\left(\frac{\Delta H_i}{H}\right)^2=\frac{1}{9}\sum_{i<j} \left(\frac{H_i-H_j}{H}\right)^2. \label{6.4}
\end{equation}
From Einstein's equations then follows
\begin{equation}
\kappa \rho=(3/2)(2-A)H^2-\Lambda, \label{6.5}
\end{equation}
and
\begin{equation}
A=C\frac{\tau^{2(3\alpha-1)}e^{-2\Phi}}{H^2}, \quad \tau=abc, \quad \Phi=2\kappa \int \eta dt, \label{6.6}
\end{equation}
where $C$ is an integration constant which is given by the present values as $C=A_0H_0^2e^{2\Phi_0}$ with the normalization $\tau(t_0)=1$.
 Insertion of eq. (\ref{6.6}) into eq. (\ref{6.5}) leads to
\begin{equation}
\kappa \rho=3H^2-(3/2)C\tau^{2(3\alpha-1)}e^{-2\Phi}-\Lambda. \label{6.7}
\end{equation}
A further insertion of eq. (\ref{6.7}) into eq. (\ref{6.3}) gives
\begin{equation}
\dot{H}=-\frac{3}{2}(1+w)H^2+\frac{3}{2}\kappa \xi H-\frac{3C}{4}(1-w)\tau^{-2(1-3\alpha)}e^{-2\Phi}+\frac{1}{2}(1+w)\Lambda. \label{6.8}
\end{equation}
Saha [2008] \cite{saha08} has investigated Bianchi type-I universe models containing fluids with different kinds of viscosity. As a first case he considered a model with vanishing shear viscosity. Then $\Phi=0$. Assuming also that there is no non-linear viscosity so that $\alpha=0$, eqs. (\ref{6.6}) and (\ref{6.7}) reduce to
\begin{equation}
A=C/(\tau H)^2=9C/\dot{\tau}^2, \label{6.9}
\end{equation}
and
\begin{equation}
\kappa \rho=3H^2-\Lambda -(3/2)C/\tau^2. \label{6.10}
\end{equation}
Saha [2008] \cite{saha08} has found a similar equation, but with positive sign before $\Lambda$ and claims that during the expansion of the universe $\tau$ increases and $\rho$ decreases. Assuming that at some stage $\tau \rightarrow \infty, \, \rho \rightarrow 0$ it follows that (with Saha's sign) $3H^2+\Lambda\rightarrow 0  $. From this Saha concludes that the universe may be infinitely large only if $\Lambda \leq 0$. It seems, however, that Saha has got the cosmological constant with wrong sign in his field equations (2.22), so the correct conclusion is that a positive cosmological constant permits the scale factor to evolve towards an infinitly large value.

\subsection{Bianchi Type-I Universe with Viscous Zel'dovich Fluid and LIVE}

Saha also considered the case where the coefficient of viscosity is inversely proportional to the Hubble parameter, $\xi H={\rm constant}=2C_\xi/3\kappa$. From eq. (\ref{6.8}) is seen that in this case the viscosity acts like a cosmological constant. Assuming further that the cosmic fluid is a Zel'dovich fluid, eq. (\ref{6.8}) takes the form (\ref{4.15}) in Sect. 4   with $a=1, b=3, c=0, d=C_\xi+\Lambda$. With the boundary conditions $H(t_0)=H_0, \tau(0)=0, \tau(t_0)=1$ it is found that the Hubble parameter, the volume scale factor and the anisotropy parameter are
\begin{equation}
H=\frac{\hat H}{3}\coth (\hat{H}t), \quad \tau=\left[\left(\frac{3H_0}{\hat H}\right)^2-1\right]^{1/2}\sinh (\hat{H}t), \quad A=A_0\left(\frac{H_0}{\tau H}\right)^2, \label{6.11}
\end{equation}
where $\hat{H}=3(C_\xi+\Lambda)$ and
$A_0$ is the present value of the anisotropy parameter. The age of the universe and the present values of the Hubble parameter are
\begin{equation}
t_0=\frac{1}{\hat H}{\rm arctanh}(\hat{H}t), \quad H_0=\sqrt{\frac{2\kappa (\rho_0+\rho_{\Lambda 0})}{3(2-A_0)}}. \label{6.12}
\end{equation}

Furthermore, Saha  considered a universe model with both bulk and shear viscosity. He assumed that the coefficient of shear viscosity is negative and proportional to the Hubble parameter, so that
\begin{equation}
(2\kappa \eta/3)=-H. \label{6.13}
\end{equation}
This gives
\[
e^{-\Phi}=\tau,
\]
so that eq. (\ref{6.7}) with $\alpha=0$ reduces to
\begin{equation}
\kappa \rho=3H^2-(3/2)C-\Lambda. \label{6.14}
\end{equation}
Hence the constant $C$ is given in terms of present quantities as $C=2H_0^2-(2\kappa/3)(\rho_0+\rho_{\Lambda 0}).$ Then eq. (\ref{6.8}) takes the form (\ref{4.15}) in Sect. 4 with
 $a=1, \, b=(3/2)(1+w), \, c=(3/2)\kappa \rho, \, d=[H_0^2-(\kappa/3)\rho_0](1-w)+\Lambda,$ and the solution is given by eqs. (\ref{4.17})-(\ref{4.20}) in Sect. 4. In the case of a Zel'dovich fluid, $b=3$ and $d=\Lambda$.

 A similar  example was considered by  Mak, Harko and Fung [1997] \cite{mak97}, who investigated a
   Bianchi type-I universe model with a viscous fluid consisting of a mixture of a Zel'dovich fluid (also called stiff matter because the velocity of sound is equal to the velocity of light) with $w=1$ and LIVE with $w=-1$. In this model the only type of viscosity was a constant bulk viscosity.
   Then eq. (\ref{6.3}) reduces to
 \begin{equation}
 \dot{H}=-3H^2+\frac{3}{2}\kappa \xi H+\Lambda, \label{6.15}
 \end{equation}
 which has the same form as eq. (\ref{4.11}) in Sect. 4 with $a=1, \, b=3, \, c=(3/2)\kappa \xi_0=(3/2)\Omega_{\xi 0}H_0, \, d=\Lambda=3\Omega_{\Lambda 0}H_0^2$. From eq. (\ref{4.13}) in Sect. 4 the Hubble parameter  then is
 \begin{equation}
 H=H_0\left[ \Omega_{\xi 0}/4+K_{\lambda \xi}\coth (3K_{\lambda \xi}H_0t)\right], \quad K_{\lambda \xi}=\sqrt{ \Omega_{\Lambda 0}+(\Omega_{\xi 0}/4)^2}. \label{6.16}
 \end{equation}
 According to eq. (\ref{4.15}) in Sect. 4 the volume scale factor normalized to unity at present time is
 \begin{equation}
 \tau=K_\xi^{1/2}e^{(3/4)\Omega_{\xi 0}H_0(t-t_0)}\sinh (3K_{\lambda \xi}H_0t), \quad K_\xi=\frac{1-\Omega_{\Lambda 0}-\Omega_{\xi 0}/2}{K_{\lambda \xi}^2}. \label{6.17}
 \end{equation}
 Equation (\ref{4.16}) in Sect. 4  gives for the age of this universe model
 \begin{equation}
 t_0=\frac{1}{3K_{\Lambda \xi}H_0}{\rm arctanh}
 \frac{K_{\Lambda \xi}}{1-\Omega_{\xi 0}/4}. \label{6.18}
 \end{equation}
 With $A (0)=2$, which is equal to the anisotropy of a Kasner universe \cite{gron85}, eq. (\ref{6.6}) gives for the anisotropy parameter
 \begin{equation}
 A=\frac{2}{e^{(1/2)(3\Omega_{\xi 0}+8\Omega_{\eta 0})H_0t}\left[ \cosh (3K_{\lambda \xi}H_0t)+\frac{\sinh (3K_{\lambda \xi}H_0t)}{\sqrt{1+16\Omega_{\lambda 0}/\Omega_{\xi 0}^2}}\right]^2}, \label{6.19}
 \end{equation}
 where $\Omega_{\eta 0}=\kappa \eta_0 /H_0$.

 The Hubble parameter and hence, from eq.~(\ref{6.7}) also the energy density of the fluid, are both infinitely large at the beginning of the cosmic evolution. As $t$ increases, the Hubble parameter and the energy density decrease and approach finite values. The universe starts from a Big Bang with a vanishing value of the volume scale factor.

 \subsection{Bianchi Type I Universe with Variable Shear and Bulk Viscosity}

 In order to consider a simple example of anisotropic universe models with variable viscosity, Mak, Harko and Fung [1997] \cite{mak97} also investigated a universe model dominated by a Zel'dovich fluid with vanishing shear viscosity and bulk viscosity proportional to the Hubble parameter $H$ averaged over the different directions, $\xi=\xi_1 H$. In this case eq.~(\ref{6.8}) reduces to
 \begin{equation}
 \dot{H}=-3\left( 1-\frac{1}{2}\kappa \xi_1\right) H^2. \label{6.20}
 \end{equation}
 The Hubble parameter and the scale factor are
 \[
 H=\frac{H_0}{1+K_{\xi 1}H_0 (t-t_0)}, \quad \tau=[1+K_{\xi 1} H_0(t-t_0)]^{3K_{\xi 1}}, \]
 \begin{equation}
 \quad K_{\xi 1}=3[1-(1/2)\kappa \xi_1]. \label{6.21}
 \end{equation}
 With the condition $A(t_0)=2$ the anisotropy parameter of this universe model is
 \begin{equation}
 A=\frac{2}{[1+K_{\xi 1}H_0(t-t_0)]^{8-\frac{9}{2}\kappa \xi_1}}. \label{6.22}
 \end{equation}
 In order that this shall represent a Big Bang universe model the volume scale factor must obey $\tau(0)=0.$ Then the age of this universe model is
 \begin{equation}
 t_0=\frac{1}{K_{\xi 1}H_0}. \label{6.23}
 \end{equation}
 In this case the Hubble parameter, volume scale factor and anisotropy parameters are
 \begin{equation}
 H=\frac{1}{K_{\xi 1}t}, \quad \tau=(K_{\xi 1}H_0t)^{3K_{\xi 1}}, \quad A=\frac{2}{(K_{\xi 1}H_0t)^{8-(9/2)\kappa \xi_1}}. \label{6.24}
 \end{equation}
 The effect of shear viscosity in such a universe may be investigated analytically by introducing a carefully chosen proportionality constant between the coefficient of shear viscosity and the Hubble parameter,
 \begin{equation}
 \eta=-\frac{3}{2}(1-3\alpha)H, \label{6.25}
 \end{equation}
 so that $\tau^{2(2\alpha-1)}e^{-2\Phi}=1$. Furthermore we assume that the coefficient of bulk viscosity depends linearly upon $H$, $\xi=\xi_0+\xi_1H$. Then eq.~(\ref{6.8}) reduces to
 \begin{equation}
 \dot{H}=-\frac{3}{2}(1+w-\kappa \xi_1)H^2+\frac{3}{2}\kappa \xi_0H-\frac{3}{4}C(1-w)+\frac{1}{2}(1+w)\Lambda. \label{6.26}
 \end{equation}

Integration with $\tau(0)=0$, $\tau(t_0)=1$ gives
 \begin{equation}
 H(t)=\frac{\kappa \xi_0}{2(1+w-\kappa \xi_1)}+\hat{H}\coth \left[\frac{3}{2}\hat{H}(1+w-\kappa \xi_1)t\right], \label{6.27}
 \end{equation}
 where
 \begin{equation}
 \hat{H}^2=\left(\frac{\kappa \xi_0}{2(1+w-\kappa \xi_1)}\right)^2+\frac{(1+w)(\Lambda/3)+(1-w)(C/2)}{1+w-\kappa \xi_1}, \label{6.28}
 \end{equation}
 and the volume factor is
 \begin{equation}
 \tau(t)=e^{\frac{3\kappa \xi_0(t-t_0)}{2(1+w-\kappa \xi_1)}}\left( \frac{\sinh \left[\frac{3}{2}\hat{H}(1+w-\kappa \xi_1)t\right]}
 {\sinh \left[ \frac{3}{2}\hat{H}(1+w-\kappa \xi_1)t_0\right]}\right)^{\frac{2}{1+w-\xi_1}}, \label{6.29}
 \end{equation}
 where
 \begin{equation}
 t_0=\frac{2}{3\hat{H}(1+w-\kappa \xi_1)}{\rm arccoth} \left[ \frac{H_0-\frac{\kappa \xi_0}{2(1+w-\kappa \xi_1)}}{\hat{H}}\right]. \label{6.30}
 \end{equation}
 In this case the anisotropy parameter is given by
 \begin{equation}
 A=C/H^2, \label{6.31}
 \end{equation}
 and the density is
 \begin{equation}
 \kappa \rho=3H^2-\Lambda -3C/2. \label{6.32}
 \end{equation}
 Here $
  C=(2/3)(3H_0^2-\kappa \rho_0-\Lambda)$, giving
  \begin{equation}
  \kappa \rho=\kappa \rho_0+3(H^2-H_0^2). \label{6.33}
  \end{equation}

 \section{Anisotropic Universe Model with Decaying Vacuum Energy}
 \setcounter{equation}{0}

 Bali, Singh and Singh [2012] \cite{bali12} have considered a related universe model with Zel'dovich fluid, $w=1$, and a decaying vacuum energy with density proportional to the Hubble parameter, $\Lambda=\alpha H$, where $\alpha$ is a positive constant. Then eq. (\ref{6.13}) in Sect. 6 reduces to
 \begin{equation}
 \dot{H}=-\frac{3}{2}(2-\kappa \xi_1)H^2+\left(\alpha+\frac{3}{2}\kappa \xi_0\right)H. \label{7.1}
 \end{equation}
 The general solution is
 \begin{equation}
 H=\frac{\beta H_0}{H_0\gamma -(H_0\gamma -\beta)e^{-\beta (t-t_0)}}, \label{7.2}
 \end{equation}
 where $\beta=\alpha+(3/2)\kappa \xi_0, \, \gamma=3-(3/2)\kappa \xi_1$ and $H_0=H(t_0)$. The initial value of the Hubble parameter is
 \begin{equation}
 H(0)=\frac{\beta H_0}{H_0\gamma -(H_0\gamma -\beta)e^{\beta t_0}}. \label{7.3}
 \end{equation}
 Considering a universe model with an initial Big Bang having $H(0)=\infty$ gives the age of the universe model in terms of the present value of the Hubble parameter,
 \begin{equation}
 t_0=-\frac{1}{\beta}\ln \left( 1- \frac{\beta}{H_0 \gamma}\right). \label{7.4}
 \end{equation}
 For this universe model the expression for the Hubble parameter reduces to
 \begin{equation}
 H=\frac{\beta}{\gamma(1-e^{-\beta t})}. \label{7.5}
 \end{equation}
 Introducing an average scale factor $R=(R_1R_2R_3)^{1/3}$ so that $H=\dot{R}/R$ and integrating with the normalization $R(t_0)=1$ we obtain
 \begin{equation}
 R=\left( \frac{e^{\beta t}-1}{e^{\beta t_0}-1}\right)^{1/\gamma}. \label{7.6}
 \end{equation}
 The decay of the density of the vacuum energy is given by
 \begin{equation}
 \Lambda=\frac{\alpha \beta}{\gamma (1-e^{-\beta t})}. \label{7.7}
 \end{equation}
 The deceleration parameter as given in eq. (\ref{2.6}) in Sect. 2 is
 \begin{equation}
 q=\gamma e^{-\beta t}-1. \label{7.8}
 \end{equation}
 At early times $q \approx 2-(3/2)\kappa \xi_1>0$ and the expansion decelerates. The deceleration is reduced by the component of the bulk viscosity proportional to the Hubble parameter. At the instant $t_1$ given by  $q(t_1)=0$, i.e. at $t_1=(1/\beta)\ln \gamma$, there is a transition from decelerated to accelerated expansion. In the limit of large times this universe model enters a de Sitter era with constant Hubble parameter
 \begin{equation}
 H(t\rightarrow \infty)=\frac{\alpha +(3/2)\kappa \xi_0}{3-(3/2)\kappa \xi_1}, \label{7.9}
 \end{equation}
 constant density of vacuum energy, and constant deceleration parameter $q(t\rightarrow \infty)=-1.$

 \section{Other Viscous Universe Models}
 \setcounter{equation}{0}

 Pradhan and Srivastava [2007] \cite{pradhan07} have investigated viscous universe models of Bianchi type V with constant viscosity coefficient, and with viscosity coefficient proportional to the density, i.e. with $\xi=\xi_0$ and $\xi=\xi_2H^2$. The line elements of these universe models have the form
 \begin{equation}
 ds^2=-\frac{1}{2}dT^2+n^2T^2 dX^2+n^{\frac{2}{n}}T^{\frac{2}{n}}e^{2X}(dY^2+dZ^2). \label{8.1}
 \end{equation}
 With this line element as a point of departure the authors calculated the time dependence of the density and the cosmological 'constant', $\Lambda(T)$. Similar universe models have been investigated by Singh and Baghel [2009] \cite{singh09}.

 Ram, Singh and Verma [2012] \cite{ram12} have investigated universe models of Bianchi type II with variable $G$ and $\Lambda$. Their solution of the field equations seem rather ad hoc, however, since they assume a specific form of the metric and calculate the time dependence of $G,\,\Lambda $ and $\xi$ from the field equations.

\section{Viscosity, Turbulence, and Big Rip/Little Rip Cosmology}
\setcounter{equation}{0}

Let us return to the observed  accelerated expansion of the universe. Current observations indicate that the equation-of-state parameter $w$ lies in the region $w=-1.04^{+0.09}_{-0.10}$;  cf. Nakamura et al. [2010] \cite{nakamura10} and Amanullah et al. [2010] \cite{amanullah10}. The case $w<-1$ (the phantom case) is the least understood region, as all the four energy conditions are violated. Although the theory is unstable from a quantum field theoretical viewpoint, it could be stable in classical cosmology. Observations indicate; cf.  Caldwell [2001] \cite{caldwell02} and  Caldwell et al. [2003] \cite{caldwell03},  that the crossing of the cosmological constant/phantom divide took place in the near past, or will occur in the near future. An essential property of most of the phantom dark energy models is the Big Rip future singularity \cite{caldwell02,caldwell03} where the scale factor becomes infinite at a finite time in the future. A softer future singularity caused by phantom
or quintessence dark energy is the so-called Type II singularity; cf. Barrow [2004] \cite{barrow04}, Nojiri and Odintsov [2004] \cite{nojiri04}, where the scale factor is finite at the Rip time. Recently, an attempt to resolve the finite-time future singularities has been proposed by introducing mild phantom models where $w$ asymptotically tends to $-1$ and where the energy density increases with time or remains constant, but where the singularity occurs in the infinite future; cf. Frampton et al. [2011, 2012] \cite{frampton11,frampton12,frampton12B}. The key point here is that if $w$ approaches $-1$ sufficiently fast, the time required for occurrence of the singularity is infinite. That is, in effect the singularity never happens.

The pioneering works on the future singularity \cite{caldwell02,caldwell03} considered the cosmic fluid to be nonviscous. Viscous Little Rip cosmology in an isotropic fluid (only bulk viscosity present) was recently worked out by Brevik et al. [2011] \cite{brevik11}. Let us give one of the characteristic results from that investigation: if the effective pressure $p_{\rm eff}$ is assumed to have the explicit form

\begin{equation}
p_\mathrm{eff}=-\rho-A\sqrt{\rho}-3\xi H
\label{9.1}
\end{equation}
with $A$ a positive constant, and if moreover the bulk viscosity $\xi$ is
assumed to satisfy the condition
\begin{equation}
3\xi H \equiv C=\mathrm{constant}\, ,
\label{9.2}
\end{equation}
then the following expression is found for the time dependent energy density
\begin{equation}
\rho(t)=\left[ \left( \frac{C}{A}+\sqrt{\rho_0}\right)
\exp (\frac{1}{2}\sqrt{3\kappa}\,At)-\frac{C}{A}\right]^2\,
\label{9.3}
\end{equation}
(recall that $\kappa=8\pi G$). Subscript zero refers to the present time. Thus an infinite time is required to reach the infinite energy density state. This is just the characteristic property of the Little Rip 'singularity'.

\subsection{A Turbulent Approach}

The simple description above, in terms of macroscopic bulk viscosity in the
fluid, cannot be considered to be satisfactory in the later stages of the development of the universe where it approaches the future singularity.
The reason is that the motion is then violent, and  a transition into
{\it turbulent} motion seems to be inevitable.
The local Reynolds number must be expected to be very high.
That brings in fact the {\it shear} viscosity back into the analysis (it is usually neglected in viscous cosmology due to the assumed spatial isotropy),  now not in a
macroscopic but in a local sense, causing  the distribution of local eddies over the wave
number spectrum. In this section we will review the cosmological turbulence theory as formulated by Brevik et al. [2012] \cite{brevik12}.

What kind of turbulence should we expect? The natural choice is that of
isotropic turbulence,
which is a topic reasonably well understood. Thus, we should expect a
Loitziankii region for
low wave numbers where the energy density varies proportionally to $k^4$;
for higher $k$ we should expect an inertial subrange characterized by the
formula
\begin{equation}
E(k)=\alpha \epsilon^{2/3}k^{-5/3}
\label{9.4}
\end{equation}
with $\alpha$ the Kolmogorov constant and $\epsilon$ the mean energy
dissipation per unit time
and unit mass; and  finally when the values of $k$ become as high as the
inverse Kolmogorov length $\eta_L$,
\begin{equation}
k \rightarrow  k_L=\frac{1}{\eta_L}=\left(\frac{\epsilon}{\nu^3}\right)^{1/4}
\label{9.5}
\end{equation}
with $\nu$ the kinematic viscosity, we enter the dissipative region where the
local Reynolds number
is of order unity and heat dissipation occurs.
In accordance with common usage we shall consider the fluid system as
quasi-stationary,
and omit the production of heat energy. In practical cases it may be useful to
combine
these elements into the useful von K\'{a}rm\'{a}n interpolation formula which
covers the whole wave number spectrum (cf., for instance, Panchev [1971] \cite{panchev71}, Brevik [1992] \cite{brevik92}, Carhart and Kostinski [1988] \cite{carhart88}.

However, the full spectral theory of isotropic turbulence will not be needed in
our first approach
to the problem. Rather, we shall in the following focus attention on how the
turbulent part
of the energy density, called $\rho_\mathrm{turb}$, can be estimated to vary
from present time $t_0$ onwards.
First, we write the effective energy density  as a sum of two terms,
\begin{equation}
\rho_\mathrm{eff} = \rho + \rho_\mathrm{turb}\, ,
\label{9.6}
\end{equation}
where $\rho$ denotes the conventional macroscopic energy density in the local
rest inertial system of the fluid.
It is natural to assume that $\rho_\mathrm{turb}$ is proportional to $\rho$
itself.
Further, we shall assume that $\rho_\mathrm{turb}$ is proportional to the
scalar expansion
$\theta={u^{\mu}}_{;\mu}=3 H$.
This because physically speaking the transition to turbulence is expected to be
more pronounced
in the violent later stages, and a proportionality to the scalar expansion is
mathematically
the most simple way in which to represent the effect. Calling the
proportionality factor $\tau$,
we can thus  write the effective energy density as
\begin{equation}
\rho_\mathrm{eff}=\rho(1+3\tau H)\, . \label{9.7}
\end{equation}
Consider next the effective pressure $p_\mathrm{eff}$. We split it into two
terms,
\begin{equation}
p_\mathrm{eff}=p+p_\mathrm{turb}\, , \label{9.8}
\end{equation}
analogously as above.
For the conventional non-turbulent quantities $p$ and $\rho$ we assume the
standard relationship
\begin{equation}
p=w\rho\, ,
\label{9.9}
\end{equation}
where $-1<w<-1/3$ in the quintessence region and $w<-1$ in the phantom region.
The question now is: How does $p_\mathrm{turb}$ depend on $\rho_\mathrm{turb}$?
There seems to be no definite physical guidance to that problem, so we shall
make
the simplest possible choice in the following, namely write
\begin{equation}
p_\mathrm{turb}=w_\mathrm{turb}\,\rho_\mathrm{turb}\, , \label{9.10}
\end{equation}
with $w_\mathrm{turb}$ a constant.

We shall consider two different possibilities for the value of
$w_\mathrm{turb}$.
The first is to put $w_\mathrm{turb}$ equal to $w$ in Eq.~(\ref{9.9}), meaning
that the
turbulent matter behaves in the same way as the non-turbulent matter as far as
the equation
of state is concerned. This option is straightforward and natural,
and is not quite trivial since $\rho_\mathrm{turb}$ and $\rho$ behave
differently,
in view of Eq.~(\ref{9.7}).
Our second option will be to assume that $w_\mathrm{turb}$ takes another,
prescribed value.
In view of the expected violent conditions near the future singularity,
it might even be natural here to chose the value $w_\mathrm{turb}=+1$, i.e.,
the Zel'dovich fluid option.

The first and the second Friedmann equations  can now be written
\begin{equation}
H^2=\frac{1}{3}\kappa \rho(1+3\tau H)\, ,
\label{9.11}
\end{equation}
\begin{equation}
\frac{2\ddot{a}}{a}+H^2=-\kappa  \rho(w+3\tau H w_\mathrm{turb}) \, .
\label{9.12}
\end{equation}
This may be compared with an earlier attempt of Brevik et al. [2011] \cite{brevik011} to introduce the turbulence in dark
energy.

Equations (\ref{9.11}) and (\ref{9.12}) determine our physical model.
Recall that its input parameters are $\{w, w_\mathrm{turb}, \tau \}$, all
assumed constant.
   From these equations we can now describe the development of the Hubble
parameter.
For convenience we introduce the quantities $\gamma$ and
$\gamma_\mathrm{turb}$, defined as
\begin{equation}
\gamma=1+w, \quad \gamma_\mathrm{turb}=1+w_\mathrm{turb}\, .
\label{9.13}
\end{equation}
We can then write the governing equation for $H$ as
\begin{equation}
(1+3\tau H)\dot{H}+\frac{3}{2}\gamma H^2+\frac{9}{2}\tau
\gamma_\mathrm{turb}H^3=0\, .
\label{9.14}
\end{equation}
This equation is in principle to be integrated from present time $t=t_0=0$
onwards, with initial value $H=H_0=\dot{a}_0/a_0$.

If $T_\mathrm{tot}^{\mu\nu}$ denotes the total energy-momentum tensor for the
cosmic fluid, we must have
\begin{equation}
{T_\mathrm{tot}^{\mu\nu}}_{;\nu}=0\, , \label{9.15}
\end{equation}
as a consequence of Einstein's equation.

In most cases studied, the expression for $T_\mathrm{tot}^{\mu\nu}$ can be
written down explicitly;
this is so for non-viscous fluids as well as with macroscopic viscous fluids.
In the present case this no longer true, however, since the turbulent energy is
produced by shear
stresses on a {\it small scale}, much less than the scale of the macroscopic
fluid equations.
That is, we are dealing with a non-closed physical system, of essentially the
same kind as encountered
in phenomenological electrodynamics in a continuous medium in special
relativity.
It implies that the source term in the energy balance equation has to be put in
by hand.

Let henceforth $T^{\mu\nu}$ refer to the non-viscous part of the fluid. We may
express the energy balance as
\begin{equation}
\dot{\rho}+3H(\rho+p)=-Q\, ,
\label{9.16}
\end{equation}
where the source term $Q$ is positive, corresponding to an energy sink for the
non-viscous fluid.
We shall put $Q$ equal to $\epsilon \rho$, where the specific energy
dissipation $\epsilon$ however
shall be taken to involve the large Hubble parameter $H$ in the later stage of
the development.
Let us assume the form
\begin{equation}
\epsilon=\epsilon_0(1+3\tau H)\, ,
\label{9.17}
\end{equation}
$\epsilon_0$ being the specific energy dissipation at present time.
This equation is seen to contain the same kind of development as assumed
before; cf. the analogous
Eq.~(\ref{9.7}). Thus, our ansatz for the energy balance reads
\begin{equation}
\dot{\rho}+3H(\rho+p) = -\rho \epsilon_0(1+3\tau H)\, .
\label{9.18}
\end{equation}

We now consider in some detail the mentioned  two different options for the value of $w_{\rm turb}$.

\subsubsection{ The Case $w_{\rm turb}=w<-1$}

This case means that the turbulent component of the fluid is regarded as a
passive ingredient
as far as the  parameter $w$ is concerned.
The time development of $\rho$ and $\rho_\mathrm{turb}$ will however be
different.
Equation~(\ref{9.14}) reduces to
\begin{equation}
\dot{H}+\frac{3}{2}\gamma H^2=0\, ,
\label{9.19}
\end{equation}
leading to
\begin{equation}
H=\frac{H_0}{Z}\, ,
\label{9.20}
\end{equation}
where we have defined
\begin{equation}
Z=1+\frac{3}{2}\gamma H_0 t
\label{9.21}
\end{equation}
(note that $w<-1$ implies $\gamma <0$).
Thus we have a Big Rip cosmology, where the future singularity time $t_s$ is
given by
\begin{equation}
t_s=\frac{2}{3|\gamma|H_0}\, .
\label{9.22}
\end{equation}
The scale factor becomes correspondingly
\begin{equation}
a=a_0Z^{2/3\gamma}\, ,
\label{9.23}
\end{equation}
and from the first Friedmann equation (\ref{9.11}) we get the non-turbulent
energy density as
\begin{equation}
\rho=\frac{3H_0^2}{\kappa^2}\,\frac{1}{Z}\,\frac{1}{Z+3\tau H_0}\, .
\label{9.24}
\end{equation}
The ratio between turbulent and non-turbulent energy becomes
\begin{equation}
\frac{\rho_\mathrm{turb}}{\rho}=3\tau H=\frac{3\tau H_0}{Z}\, .
\label{9.25}
\end{equation}
It is of main interest to consider the behavior near $t_s$. As $Z=1-t/t_s$ we
see that
\begin{equation}
H \sim \frac{1}{t_s-t}\, , \quad a \sim \frac{1}{(t_s-t)^{2/3|\gamma|}}\, ,
\label{9.26}
\end{equation}
\begin{equation}
\rho \sim \frac{1}{t_s-t}, \quad \frac{\rho_\mathrm{turb}}{\rho} \sim
\frac{1}{t_s-t}\, .
\label{9.27}
\end{equation}
Notice the difference from conventional cosmology: the behavior of $H$ and $a$
near the singularity
is as usual, while the singularity of $\rho$ has become {\it weakened}.
The reason for this is, of course, the non-vanishing value of the parameter
$\tau$.
Moreover, as $t \rightarrow t_s$ all the non-turbulent energy has been
converted into turbulent energy.
   From a physical point of view, this is just as we would expect.

\subsubsection{The Case $w<-1, w_{\rm turb}>-1$}

This case is thermodynamically quite different from the preceding one as the
turbulent component of the fluid is no longer a passive ingredient.
We have now $\gamma_\mathrm{turb}=1+w_\mathrm{turb} >0$,
which means that we  cover  the  region $-1<w_\mathrm{turb}<0$
also.
In the latter region, the turbulent
contribution to the pressure is still negative as above, while if
$w_\mathrm{turb}>0$
the turbulent pressure becomes {\it positive}, just as in ordinary
hydrodynamical turbulence.

The governing equation (\ref{9.14}), written as
\begin{equation}
(1+3\tau H)\dot{H}=\frac{3}{2}H^2(|\gamma|-3\tau \gamma_\mathrm{turb}H)\, ,
\label{9.28}
\end{equation}
tells us that at the present time $t=0$ the condition
\begin{equation}
|\gamma| > 3\tau \gamma_\mathrm{turb}H_0
\label{9.29}
\end{equation}
must hold, because at the present time the turbulent part is regarded as
unimportant.
This corresponds to the inequality $\dot{H}>0$.

Equation~(\ref{9.28}) can be integrated to give $t$ as a function of $H$,
\begin{equation}
t=\frac{2}{3|\gamma|}\left(\frac{1}{H_0}-\frac{1}{H}\right)-\frac{2\tau}{|\gamma|}
\left( 1+\frac{\gamma_\mathrm{turb}}{|\gamma|}\right)
\ln \left[ \frac{|\gamma|-3\tau \gamma_\mathrm{turb}H}{|\gamma|-3\tau
\gamma_\mathrm{turb}H_0}\frac{H_0}{H}\right] \, .
\label{9.30}
\end{equation}
A striking property of this expression is that it describes a {\it Little Rip}
scenario.
As $t\rightarrow \infty$, the Hubble parameter reaches a finite critical value
\begin{equation}
H_\mathrm{crit}=\frac{1}{3\tau}\frac{|\gamma|}{\gamma_\mathrm{turb}}\, .
\label{9.31}
\end{equation}
The physical role of $\gamma_\mathrm{turb}$ is thus to postpone and weaken
the development towards the future singularity.

A natural choice for the EoS parameter $w_\mathrm{turb}$ in the vicinity of the
singularity,
in view of the violent motions expected, would be
\begin{equation}
w_\mathrm{turb}=+1\, ,
\label{9.32}
\end{equation}
that means, a Zel'dovich fluid. This is an extreme case,
where the velocity of sound equals the velocity of light.

\subsection{A One-Component Dark Fluid}

We now turn to an approach that is quite different from
the one above,
namely to consider the cosmic fluid as a {\it one-component} fluid.
Thus the distinction between a non-turbulent and a turbulent fluid component is
avoided altogether.
This new approach is actually more close to the usual picture in hydrodynamics,
where
a fluid is known to shift suddenly from a laminar to a turbulent state.

Consider the following picture: the universe starts from present time $t=0$ as
an ordinary
viscous fluid with a bulk viscosity  $\xi$, and develops
according to the Friedmann equations. We assume as before that the EoS
parameter $w<-1$,
meaning that the universe develops in the viscous era towards a future
singularity.
Before this happens, however, at some instant $t=t_*$, we assume that there is
a sudden transition of the whole fluid into a turbulent state, after which the
EoS parameter is $w_\mathrm{turb}$
and the pressure accordingly $p_\mathrm{turb}=w_\mathrm{turb}
\,\rho_\mathrm{turb}$.
As before we assume that $w_\mathrm{turb} >-1$, and for simplicity we take
$\xi$, as well as $w$ and $w_\mathrm{turb}$, to be constants.
One may ask: What is the resulting behavior of the fluid, especially at later
stages?

The problem can easily be solved, making use of the condition that the density
of the fluid has to be continuous at $t=t_*$. One gets; cf. Brevik and Gorbunova [2005]  \cite{brevik05}, Brevik et al. [2010] \cite{brevik10}
\begin{equation}
H=\frac{H_0\,e^{t/t_c}}{1-\frac{3}{2}|\gamma|H_0t_c(e^{t/t_c}-1)}\, ,
\label{9.33}
\end{equation}
\begin{equation}
a=\frac{a_0}{\left[1-\frac{3}{2}|\gamma|H_0t_c(e^{t/t_c}-1)
\right]^{2/3|\gamma|}}\, ,
\label{9.34}
\end{equation}
\begin{equation}
\rho=\frac{\rho_0\, e^{2t/t_c}}{\left[1-\frac{3}{2}|\gamma|H_0t_c(e^{t/t_c}-1)
\right]^2}\, ,
\label{9.35}
\end{equation}
where $t_c$ is the `viscosity time'
\begin{equation}
t_c=\left(\frac{3}{2}\kappa \xi \right)^{-1}\, .
\label{9.36}
\end{equation}
The values $H_*, a_*, \rho_*$ at $t=t_*$ are thereby known.

In the turbulent era $t>t_*$ we can make use of the same expressions
(\ref{9.33}) - (\ref{9.35}) as above, only with substitutions
$t_c \rightarrow \infty~(\xi \rightarrow 0)$, $t \rightarrow t-t_*,~ w
\rightarrow w_\mathrm{turb}$,
$H_0 \rightarrow H_*$, $a_0 \rightarrow a_*$, $\rho_0 \rightarrow \rho_*$. Thus
\begin{equation}
H=\frac{H_*}{1+\frac{3}{2}\gamma_\mathrm{turb}H_*(t-t_*)}\, ,
\label{9.37}
\end{equation}
\begin{equation}
a=\frac{a_*}{\left[ 1+\frac{3}{2}\gamma_\mathrm{turb}H_*(t-t_*)
\right]^{2/3\gamma_\mathrm{turb}}}\, ,
\label{9.38}
\end{equation}
\begin{equation}
\rho=\frac{\rho_*}{\left[ 1+\frac{3}{2}\gamma_\mathrm{turb}H_*(t-t_*)
\right]^{2}}
\label{9.39}
\end{equation}
(recall that $\gamma_\mathrm{turb} >0$).
Thus the density $\rho$, at first increasing with increasing $t$ according to
Eq.~(\ref{9.35}),
decreases again once the turbulent era has been entered, and goes smoothly
to zero as $t^{-2}$ when $t\rightarrow \infty$.
In this way the transition to turbulence protects the universe from entering
the future singularity.

It should be noted that whereas the density is continuous at $t=t_*$ the
pressure is not:
In the laminar era $p_*= w \rho_* <0$, while in the turbulent era $p_*
=w_\mathrm{turb}\,\rho_* $
will even be positive, if $w_\mathrm{turb}>0$.
Thus, we demonstrated the possible role of turbulence to protect the universe
from the future singularity. In the same fashion, one can consider its role in
protecting the universe from the Rip, i.e. the disintegration of bound structures.

\section{On Causal Cosmology}
\setcounter{equation}{0}

Dissipative dark energy models, in which the negative pressure responsible for the current acceleration is an effective pressure, is a nonequilibrium phenomenon. A viscous pressure can in principle play the role of an agent that drives the present acceleration of the universe; cf. Zimdahl et al. [2001] \cite{zimdahl01} and Balakin et al. [2003] \cite{balakin03}. Actually, the possibility of a viscosity dominated late epoch of the universe with accelerated expansion was discussed already in 1987 by Padmanabhan and Chitre \cite{padmanabhan87}.

Traditionally, the nonequilibrium thermodynamical processes have been formulated in the context of the classic theories of Eckart [1940] \cite{eckart40} and Landau and Lifshitz [1958] \cite{landau58}. Due to works of M\"{u}ller [1967] \cite{muller67}, Israel [1976] \cite{israel76B}, Israel and Stewart [1979] \cite{israel79} and others it became clear that Eckart-type formulations suffer from serious drawbacks concerning causality and stability. The reason for this is obviously the restriction of these early theories to first-order deviations from equilibrium. If one includes second order deviations, the problems should most likely disappear. Cosmological implications of second order theories were first considered by Belinskii et al. [1979] \cite{belinskii79} and  Pavon et al. [1982] \cite{pavon82}.

Whereas the influence from a bulk viscosity on the background expansion of the universe is widely investigated in the literature, the perturbative analysis of the viscous cosmological models is not so widely addressed. A common characteristic of  investigations of the latter kind is the analysis of a fluid whose equilibrium pressure is small compared with the bulk viscous pressure. Another common feature is that the dissipation is described within Eckart's theory. It is noteworthy that there are choices for the bulk viscosity parameters that give predictions for the matter power spectrum in agreement with observations; cf. Hipolito-Ricaldi et al. [2009] \cite{hipolito09}, and Hipolito-Ricaldi et al. [2010] \cite{hipolito10}.

The number of research papers in this area is large, so that we will in the follow focus essentially on the recent extensive paper of Piattella et al. [2011] \cite{piattella11}. That paper relies on cosmological dynamics as realized by one single imperfect fluid with bulk viscosity and vanishing equilibrium pressure. An analysis is  carried out showing how this conventional picture is changed when one takes into account the causal M\"{u}ller-Israel-Stewart theory, instead of Eckart's theory. The gravitational potential is evaluated, and compared with its $\Lambda$CDM counterpart. As is usual, it is assumed that the standard $\Lambda$CDM model in effect reproduces the observations.

First, let us write the viscous contribution to the total pressure as
\begin{equation}
\Pi=-\theta \xi, \label{10.1}
\end{equation}
where $\theta=3H$ is the scalar expansion and $\xi$ the bulk viscosity. This is the Eckart theory \cite{eckart40}. By contrast, in the M\"{u}ller-Israel-Stewart (MIS) theory one has
\begin{equation}
\tau_{\rm rel}\dot{\Pi}+\Pi=-\theta \xi -\frac{1}{2}\Pi \left[\theta -\frac{(\xi/\tau_{\rm rel})^{\bf .}}{(\xi/\tau_{rel})}-\frac{\dot{T}}{T}\right], \label{10.2}
\end{equation}
where $\tau_{\rm rel}$ is a relaxation time and $T$ is the temperature. Overdot means time derivative. This equation is often emplyed in a truncated form, as
\begin{equation}
\tau_{\rm rel}\dot{\Pi}+\Pi=-\theta \xi. \label{10.3}
\end{equation}
Taking $\Pi$ to be subject to causal thermodynamics implies that the relaxation time $\tau_{\rm rel}$ is introduced as an additional parameter. Also, a propagation velocity for viscous perturbations is introduced that is different from the adiabatic sound velocity.

The variation of the temperature in Eq.~(\ref{10.2}) is found by means of Gibbs' integrability condition; cf. Maartens [1996] \cite{maartens96}
\begin{equation}
n\frac{\partial T}{\partial n}+(\rho+p)\frac{\partial T}{\partial \rho}=T\frac{\partial p}{\partial \rho}, \label{10.4}
\end{equation}
$n$ being the particle number density. From this, by integration,
\begin{equation}
\frac{\dot T}{T}=-\theta \left[ \frac{\partial p}{\partial \rho}+\frac{\Pi}{T}\frac{\partial T}{\partial \rho}\right]. \label{10.5}
\end{equation}
Limiting ourselves to barotropic conditions, we have $T=T(\rho)$ and $p=p(\rho)$. With $c_s=\sqrt {\partial p/\partial \rho}=\sqrt{ dp/d\rho}$  as the adiabatic sound velocity (in natural units), Eqs.~(\ref{10.4}) and (\ref{10.5}) simplify to
\begin{equation}
\frac{1}{T}\frac{dT}{d\rho}=\frac{c_s^2}{\rho +p}, \label{10.6}
\end{equation}
\begin{equation}
\frac{\dot T}{T}=-\theta c_s^2\left( 1+\frac{\Pi}{\rho+p}\right). \label{10.7}
\end{equation}
The MIS transport equation then gets the simplified form
\begin{equation}
\tau_{\rm rel}\dot{\Pi}+\Pi=-\theta \xi -\frac{1}{2}\tau_{\rm rel}\Pi \left[ \theta -\frac{(\xi/\tau_{\rm rel})^.}{(\xi/\tau_{\rm rel})}+\theta c_s^2\left( 1+\frac{\Pi}{\rho+p}\right) \right]. \label{10.8}
\end{equation}
The coefficients $\tau_{\rm rel}$ and $\xi$ are in general functions of time, but are not completely arbitrary as they are related to the sound velocity; cf. Hiscock and Lindblom [1983] \cite{hiscock83} and Maartens [1996] \cite{maartens96},
\begin{equation}
c_b^2=\frac{\xi}{(\rho+p)\tau_{\rm rel}}. \label{10.9}
\end{equation}
Piattella et al. [2011] \cite{piattella11} now consider the perturbative dynamics of the viscous fluid, making use of the gauge-invariant formalism due to Bardeen. The FRW metric is perturbed in the form
\begin{equation}
ds^2=a(\eta)^2[-(1+2\Phi)d\eta^2+(1-2\Phi)d{\bf x}^2], \label{10.10}
\end{equation}
where  $\Phi$ is the gravitational potential. This potential is calculated in the Eckart, as well as in the MIS, theory. Thus the perturbed bulk viscous contribution $\delta \Pi$ depends on which underlying theory is chosen.

We shall not here go into detail about the perturbation formalism, but mention that the following ansatz is made for the viscous pressure:
\begin{equation}
\frac{\Pi}{\rho}=\rho_0\left[ \mu+(1-\mu)a^{-3n/2}\right]^{2/n}, \label{10.11}
\end{equation}
with $\mu$ and $n$ constants. This is the equation of state for a Chaplygin gas. For $n=2$ the constant pressure case is recovered. From the energy conservation equation,\ one gets, with $a_0=1$ at present time,
\begin{equation}
\rho=\rho_0\left[ \mu+(1-\mu)a^{-3n/2}\right]^{2/n}. \label{10.12}
\end{equation}
We also mention that for the truncated MIS theory, for which one has
\begin{equation}
\dot{\Pi}+\frac{1}{\tau_{\rm rel}}\Pi=-\frac{3}{\tau_{\rm rel}}H\xi, \label{10.13}
\end{equation}
the following expression is found for the relaxation time as a function of density:
\begin{equation}
\tau_{\rm rel}(\rho)=\frac{1}{\sqrt {3\kappa}}\,\frac{2\mu (\rho/\rho_0)^{-n/2}}{2c_b^2+\mu(2-n)(\rho/\rho_0)^{-n/2}[1-\mu(\rho/\rho_0)^{-n/2}]}. \label{10.14}
\end{equation}

Piattella et al. [2011] \cite{piattella11} compare their numerical solutions for the gravitational potentials with the results for the standard $\Lambda$CDM case. Eckart's and the full causal theory seem to be disfavored, whereas the truncated theory, perhaps surprisingly, turns out to be more promising: it leads to results similar to those of the $\Lambda$CDM model for a bulk viscous speed in the interval $10^{-11} \ll c_b^2 \leq 10^{-8}$.

There are, as mentioned, a large number of papers dealing with viscous cosmology. We may mention the review article of Maartens [1995] \cite{maartens95}, already referred to earlier, and also papers of Pavon et al. [1991] \cite{pavon91}, and Chimento and Jakubi [2012] \cite{chimento12}.

\section{Summary}

 Inclusion of viscosity - a most natural way to go from the standpoint of general fluid mechanics -
 may be of importance both in the initial and in the final stages of the universe. We have surveyed generalized cosmology theories  with the inclusion of bulk and shear viscosities,  having  the property that they tend to smooth out eventual anisotropies in the universe. Turbulence is also a phenomenon which one should expect to be of importance, not least so in the final stages close to the future singularities where the fluid motion is expected to be vigorous.  As we have pointed out in Section 9, one possible role of the turbulence in a one-component fluid model is to protect the universe from the future singularity.

 Causal cosmology, constructed so as to be better in accordance with basic physics than the 1940 classic theory of Eckart \cite{eckart40}, ought to attract interest, although the causal formalism becomes mathematically somewhat  more involved than the classic one. As mentioned in Section 10, the truncated version of the causal formalism seems, perhaps surprisingly,  to be the variant that most closely gives results in accordance with the standard $\Lambda$CDM formalism, and thus with the observations.


\end{document}